\documentclass[runningheads,final]{llncs}

\usepackage[utf8]{inputenc}
\usepackage[T1]{fontenc}
\usepackage{url}
\usepackage[hidelinks]{hyperref}
\hypersetup{bookmarksdepth=2,bookmarksopen,bookmarksnumbered,breaklinks}
\usepackage{breakurl}
\usepackage{graphicx}

\usepackage{pifont}

\usepackage{xcolor}

\usepackage{mdframed}
\mdfsetup{hidealllines=true,backgroundcolor=blue!10, skipabove=3pt, skipbelow=3pt}

\usepackage[obeyFinal]{todonotes}

\usepackage{comment}
\usepackage[verbose]{wrapfig}
\usepackage{csquotes}

\usepackage{enumitem}
\setlist[description]{leftmargin=\parindent,labelindent=0.5\parindent}

\usepackage{subfiles}

\usepackage{amsmath}
\usepackage{amssymb}
\usepackage{stmaryrd}
\let\emptyset\varnothing
\usepackage{mathtools}
\usepackage[
    group-digits=integer, group-minimum-digits=4, % group digits by thousands
    group-separator={,},
    free-standing-units, unit-optional-argument, % easier input of numbers with units
]{siunitx}

\usepackage{wasysym}
\usepackage{stackengine}

\usepackage[numbers,sort&compress]{natbib}
\usepackage[capitalize]{cleveref}

\usepackage{inconsolata}
\usepackage[frozencache,cachedir=minted-cache]{minted}
\setminted{
	autogobble=true,
	tabsize=4,
	texcomments=true,
    linenos=true,
    numbersep=6pt
}

\usepackage{subcaption}
\usepackage{pgfplots}
\usepgfplotslibrary{groupplots}

\pgfplotsset{
    compat=1.9,
    log ticks with fixed point, % no scientific notation in plots
    table/col sep=tab, % only tabs are column separators
    unbounded coords=jump, % better have skips in a plot than appear to be interpolating
    filter discard warning=false, % Don't complain about empty cells
}

\usepackage{booktabs}
\usepackage{colortbl}
\usepackage{tablefootnote}
\usepackage{multirow}

\usepackage{orcidlink}

\newcommand{\widen}{\mathbin{\nabla}}
\newcommand{\narrow}{\mathbin{\mathrm{\Delta}}}
\newcommand{\dualnarrow}{\ensurestackMath{\mathbin{{\stackinset{c}{0ex}{c}{1.1ex}{\scriptstyle\sim}{\mathrm{\Delta}}}}}}

\newcommand{\sqleq}{\sqsubseteq}

\newcommand{\sem}[1]{\llbracket#1\rrbracket}
\newcommand{\sems}[2][]{#1\sem{#2}^\sharp}

\DeclareMathOperator{\havoc}{havoc}
\DeclareMathOperator{\assume}{assume}
\DeclareMathOperator{\unassume}{unassume}

\DeclareMathOperator{\explode}{explode}

\newcommand{\sqcupstar}{\ensurestackMath{\mathbin{{\stackinset{c}{0ex}{c}{0ex}{\scriptstyle\star}{\sqcup}}}}}
\newcommand{\sqleqstar}{\ensurestackMath{\mathrel{{\stackinset{c}{0ex}{c}{0.2ex}{\scriptstyle\star}{\sqleq}}}}}
\DeclareMathOperator{\dom}{dom}

\newcommand{\sqcapbullet}{\ensurestackMath{\mathbin{{\stackinset{c}{0ex}{c}{0ex}{\scriptstyle\bullet}{\sqcap}}}}}

\makeatletter % \moverlay by D. Arsenau
\def\moverlay{\mathpalette\mov@rlay}
\def\mov@rlay#1#2{\leavevmode\vtop{%
        \baselineskip\z@skip \lineskiplimit-\maxdimen
        \ialign{\hfil$\m@th#1##$\hfil\cr#2\crcr}}}

\newcommand{\charfusion}[3][\mathord]{
    #1{\ifx#1\mathop\vphantom{#2}\fi
        \mathpalette\mov@rlay{#2\cr#3}
    }
    \ifx#1\mathop\expandafter\displaylimits\fi}
\makeatother
\newcommand{\bigsqcapbullet}{\charfusion[\mathop]{\bigsqcap}{\bullet}}
\newcommand{\bigsqcupstar}{\charfusion[\mathop]{\bigsqcup}{\star}}

\newcommand{\restrict}[1]{\vert_{#1}}

\title{Correctness Witness Validation by \\ Abstract Interpretation}

\author{
    Simmo Saan\inst{1}\orcidlink{0000-0003-4553-1350} \and
    Michael Schwarz\inst{2}\orcidlink{0000-0002-9828-0308} \and
    Julian Erhard\inst{2}\orcidlink{0000-0002-1729-3925} \and
    Helmut Seidl\inst{2}\orcidlink{0000-0002-2135-1593} \and\\
    Sarah Tilscher\inst{2}\orcidlink{0009-0009-9644-7475} \and
    Vesal Vojdani\inst{1}\orcidlink{0000-0003-4336-7980}
}
\authorrunning{S. Saan et al.}

\institute{
    University of Tartu, Tartu, Estonia \\
    \email{\{simmo.saan, vesal.vojdani\}@ut.ee} \and
    Technische Universit\"at M\"unchen, Garching, Germany \\
    \email{\{m.schwarz, julian.erhard, helmut.seidl, sarah.tilscher\}@tum.de}
}

\begin{document}

\maketitle

\begin{abstract}
Witnesses record automated program analysis results and make them exchangeable.
To validate correctness witnesses through abstract interpretation,
we introduce a novel abstract operation unassume.
This operator incorporates witness invariants into the abstract program state.
Given suitable invariants, the unassume operation can accelerate fixpoint convergence and yield more precise results.
We demonstrate the feasibility of this approach by augmenting an abstract interpreter with unassume operators
and evaluating the impact of incorporating witnesses on performance and precision.
Using manually crafted witnesses, we can confirm verification results for multi-threaded programs
with a reduction in effort ranging from 7\% to 47\% in CPU time.
More intriguingly, we discover that using witnesses from model checkers can guide our analyzer to verify program properties that it could not verify on its own.

\end{abstract}

\section{Introduction}

Automated software verifiers can be faulty and may produce incorrect results.
To increase trust in their verdicts,
verifiers may produce \emph{witnesses} that expose their reasoning.
Such proof objects allow independent validators to confirm analysis results.
The use of witnesses as a standardized way to communicate between different automated software verifiers was pioneered by~\citet{Beyer2015}.
They introduced an analyzer-agnostic automaton-based format for explaining property violations.
The witness automaton guides the validator towards a feasible counterexample.
This witness format was later extended to explain program correctness using invariants~\cite{Beyer2016}.
Witnesses form a cornerstone of the annual software verification competition SV-COMP~\cite{Beyer2023}
and have played a key role in the emergence of Cooperative Verification~\cite{CoVeriTeam2022,HaltermannWehrheim2022},
where independent verifiers collaborate by exchanging witnesses~\cite{BeyerWahrheim2020}.

This paper aims to show how \emph{correctness witnesses} can be validated using abstract interpretation.
Existing validators are based on model checking~\cite{Beyer_2022}, test execution~\cite{Beyer_2018}, interpretation~\cite{_vejda_2020,Ayaziov__2022}, or SMT-based verification~\cite{Ponce_de_Le_n_2022};
whereas, validators for correctness witnesses at SV-COMP 2023 were all based on model checking~\cite{Beyer2016,Beyer2023}.
In that year's competition report, the long-time organizer Dirk Beyer highlighted the scarcity of validators --- which leaves many verification outcomes without independent confirmation --- as a \enquote{remarkable gap in software-verification research}~\cite{Beyer2023}.
Abstract interpretation, originally proposed by~\citet{CousotC77}, has proven successful in the efficient verification of large real-world software~\cite{Cousot2009,BaudinBBCKKMPPS21} and multi-threaded programs~\cite{Mine2012,Monat_2017,Farzan12,Schwarz_2021,SchwarzSSEV23}.
To complement existing validators, we propose enhancing the framework of abstract interpretation to incorporate invariants from witnesses.

For communication across technological boundaries, correctness witnesses must be restricted to invariants that do not expose internal abstractions of tools.
For example, as each tool may abstract dynamically allocated memory differently, invariants about the content of such memory may only be expressed indirectly, e.g., via C invariants such as  $\texttt{*p} \geq 0$.
The challenge is how to incorporate such tool-independent invariants into an abstract interpreter.
A key technical contribution of this paper are techniques to incorporate witness invariants, given as expressions, into abstract domains without relying on those invariants to actually hold.
This differs from existing work on witnesses for abstract interpretation (detailed in \cref{sec:related-work}), which does not allow for or aim at the exchange of witnesses across tool boundaries.

Our solution is to introduce a new abstract operation \emph{unassume}.
This operator can be used to selectively inject imprecision (hence the name) to speed up fixpoint computation.
Suitably increasing abstract values during fixpoint computation can also
improve the precision of an existing analysis, most notably due to the non-monotonicity of widening operators.
The following example illustrates both the speedup and increase in precision.\\[-5mm]
\begin{wrapfigure}[6]{r}{0.25\textwidth}
    \raggedleft
    \begin{subfigure}{0.23\columnwidth}
        \begin{minted}{c}
            int x = 40;
            while (x != 0) {
                x--;
            }
        \end{minted}
    \end{subfigure}
\end{wrapfigure}
\addtocounter{figure}{-1} % Exclude captionless wrapfigure from figure numbering
\begin{example}
    \label{ex:illustrative}
    Consider the example (shown right) from \citet{Mine2017}.
    An abstract interpreter using interval abstraction first reaches the loop head on line 2 with the interval~$[40, 40]$ for \texttt{x}.
    After one iteration, the loop head is reached with $[39, 39]$, so the abstract value at that point is~$[39, 40]$.
    To accelerate fixpoint iteration for termination, standard interval widening is applied, which abandons the unstable lower bound, resulting in $[-\infty, 40]$.
    Another iteration with this interval reaches the loop head with~$[-\infty, 39]$, which is subsumed by the previous abstract value.
    Subsequent standard interval narrowing cannot improve the inferred invariant $[-\infty, 40]$ at the loop head; therefore, the analysis fails to establish a lower bound for \texttt{x}.

    Now, suppose that a witness provides the invariant $0 \leq \texttt{x} \leq 40$ for the loop head at line 2.
    When guiding the fixpoint iteration with this witness, the loop head is again first reached with the singleton interval $[40, 40]$. Using the provided invariant, the unassume operator relaxes the lower bound of the interval:
    \[
    \sems{\unassume(0 \leq \texttt{x} \leq 40)}\, \{\texttt{x} \mapsto [40, 40]\} = \{\texttt{x} \mapsto [0, 40]\}.
    \]
    After one iteration, the loop head is now reached with $[0, 39]$, which makes the abstract value at that point $[0, 40]$.
    Thus, a fixpoint is reached without the need for widening or narrowing, and the stronger invariant $[0, 40]$ is confirmed.
    This demonstrates how the same analysis, when guided, can validate an invariant that it could not infer on its own.
    A well-chosen witness invariant can prevent precision loss during widening that cannot be recovered by narrowing, serving as a proxy for providing known widening thresholds.
    Additionally, the witness-guided analysis required fewer steps (transfer function evaluations and fixpoint iterations).
    Using the same invariant as a widening threshold does not yield such speedup.
\end{example}

After introducing relevant terms (\cref{sec:preliminaries}) and discussing the shortcomings of intuitive approaches to validate witnesses with abstract interpretation (\cref{sec:initialization-based}), the paper presents the following main contributions:
\begin{itemize}
    \item a specification of the unassume operator, and a general realization for relational abstract domains using dual-narrowing (\cref{sec:unassume});
    \item an efficient algorithm for unassuming in non-relational abstract domains (\cref{sec:indirect-unassume}), with generalization to pointer variables (\cref{app:pointers});
    \item an implementation of an abstract-interpretation--based witness validator, which is evaluated using hand-crafted invariants for multi-threaded programs and invariants produced by state-of-the-art model checkers for intricate literature examples (\cref{sec:evaluation}).
\end{itemize}
Our evaluation results provide practical evidence of the unassumed witness invariants making the analysis faster and more precise.

\section{Preliminaries}
\label{sec:preliminaries}
In the following,
we formally introduce the notion of location-based correctness witnesses,
subsequently referred to simply as \emph{witnesses},
and recall the basics of abstract interpretation.

\subsection{Witnesses}
\label{sec:witnesses}
Following the refined definitions of \citet{Beyer2022}, a correctness witness should contain hints for the proof
of program correctness.
Witness automata~\cite{Beyer2016} are a powerful way to provide such hints,
but \citet{JanNotes} has observed that their control-flow semantics are ambiguous, impairing interoperability.
In practice, however, invariants per program location are often sufficient~\cite{Beyer2016,Beyer_2022a},
which has led the SV-COMP community to adopt them~\cite{TacasMeeting}.
Therefore, we consider here correctness witnesses consisting of location-based invariants.

Let $\mathcal{N}$ denote the set of program locations.
For clarity of exposition, we consider a fixed set $\mathcal{V}$ of program variables.
Invariants from some language $\mathcal{E}$, which we do not fix, are used to specify properties of the program executions reaching a particular program location.
We assume that there is a trivial invariant $e_{\textsf{true}} \in \mathcal{E}$ that always holds.
Since the goal is to exchange invariants between tools, the choice of an invariant language involves a trade-off:
\begin{enumerate}
    \item \citet{Beyer2016} use boolean-valued side-effect--free C expressions for their invariants.
    The chief advantage is its conceptual simplicity: the semantics of such assertions is well-known,
    and analyzers already come with the necessary facilities to manipulate these expressions, as they appear in the analyzed program.
    In C expressions, pointers allow exchanging information also about the heap between verifiers.
    Nevertheless, the expressivity of such invariants is limited, especially regarding more complex data structures.
    \item ACSL~\cite{ACSL} has more expressive power than plain C expressions by offering quantification, memory predicates, etc.
    On the downside, considerably fewer analyzers support it, limiting exchange possibilities.
    \item Custom invariant languages can be arbitrarily expressive, at the cost of restricting communication to few similar tools whose re-verification to boost.
\end{enumerate}
With these notions, we introduce the definitions of a witness, its validation,
and witness-guided verification.

\begin{definition}
A \emph{witness} for a safety property $\Phi$ of a program $P$ is a tuple $(W, P, \Phi)$, where $W$ is a total mapping $W: \mathcal{N}\to\mathcal{E}$ from the program locations of $P$ to invariants from $\mathcal{E}$.
\end{definition}

The textual format in which witnesses are exchanged is not required to provide invariants for all program locations ---
we implicitly assume that if the witness contains no information for some program location,
then this location is mapped to $e_{\textsf{true}}$.
If the invariant language contains a contradictory expression $e_{\textsf{false}} \in \mathcal{E}$ that never holds, then it can be used to convey unreachability of a program location.
This notion of a witness is generic and can be instantiated to different programming and invariant languages.
For our examples, we use an invariant language of arithmetic and boolean expressions, enriched with basic pointers, address-taking (\texttt{\&}) and dereferencing (\texttt{*}).
The pointer constructs pose practical challenges, as will become apparent in subsequent sections.

\begin{definition}
    A witness $(W, P, \Phi)$ is \emph{valid} if
    \begin{enumerate}
        \item $P$ satisfies the property $\Phi$;
        \item whenever the execution of $P$ reaches the location $n \in \mathcal{N}$, the invariant $W\,n$ holds.
    \end{enumerate}
\end{definition}

\noindent
A \emph{witness validator} attempts to prove that a witness is valid; specifically,
it tries to recreate the proof that the program satisfies the property $\Phi$,
and checks that the witness makes only true claims about the program.
However, the validation track at SV-COMP 2023 scored participants according to a limited form of validation which only confirms the first condition~\cite{Beyer2023}.

A \emph{witness-guided verifier} uses the witness as guidance towards the verification of $\Phi$.
A sound verifier can perform this task without assuming the witness invariants to be true;
therefore, it qualifies as a sound validator of the first condition.
It may additionally verify the invariants in $W$ to perform full witness validation.

\subsection{Abstract Interpretation}
\label{sec:abstract-interpretation}

We rely on the framework of abstract interpretation as introduced by \citet{CousotC77,Cousot1992frameworks}, and briefly recall relevant notions here.
Let $S$ denote the set of all concrete program states.
Its subsets are abstracted by an \emph{abstract domain}~$\mathbb{D}$ satisfying the following properties~\cite{Mine2017}:
\begin{itemize}
    \item a \emph{partial order} $\sqleq$, modeling the relative precision of abstract states;
    \item a monotonic \emph{concretization} function $\gamma: \mathbb{D} \to 2^S$, mapping an abstract element to the set of concrete states it represents;
    \item a least element $\bot$, representing unreachability, i.e. $\gamma\,\bot = \emptyset$;
    \item a greatest element $\top$, representing triviality, i.e. $\gamma\,\top = S$;
    \item sound abstractions \emph{join} ($\sqcup$) and \emph{meet} ($\sqcap$) of $\cup$ and $\cap$ on $S$, respectively, i.e. $\gamma\,x\cup\gamma\,y\subseteq\gamma\,(x\sqcup y)$ and $\gamma\,x\cap\gamma\,y\subseteq\gamma\,(x\sqcap y)$ for all $x,y\in\mathbb{D}$;
    \item a \emph{widening} ($\widen$) operator, computing upper bounds that ensure termination in abstract domains with infinite ascending chains, i.e. $x\sqleq x\widen y$ and $y\sqleq x\widen y$ for all $x,y\in\mathbb{D}$, and for every sequence~$(y_i)_{i \in \mathbb{N}}$ from $\mathbb{D}$, the sequence $(x_i)_{i \in \mathbb{N}}$ defined by $x_0 = y_0$, ${x_{i+1} = x_i\widen y_{i+1}}$ is ultimately stable;
    \item a \emph{narrowing} ($\narrow$) operator, recovering some precision given up by widening, i.e. $x\sqcap y\sqleq x\narrow y \sqleq x$ for all $x,y\in\mathbb{D}$, and for every sequence $(y_i)_{i \in \mathbb{N}}$ from~$\mathbb{D}$, the sequence $(x_i)_{i \in \mathbb{N}}$ defined by $x_0 = y_0$, $x_{i+1} = x_i\narrow y_{i+1}$ is ultimately stable.
\end{itemize}
An abstract interpreter uses an abstract domain $\mathbb{D}$ and sound abstractions ${\sems{s}: \mathbb{D} \to \mathbb{D}}$ of primitive statements $s$ to model the abstract semantics of a program.
Fixpoint iteration (potentially with widening and narrowing) is used to compute for each program location an abstract state, which represents a superset of all reaching concrete program states.
The resulting abstract states may be used to check whether the program satisfies a given safety property.

For \emph{validating} a witness by abstract interpretation, we assume that the analyzer provides us with a mapping $\sigma: \mathcal{N} \to \mathbb{D}$ from locations to abstract values.
A witness $(W, P, \Phi)$ is validated by the abstract interpreter, if
\begin{enumerate}
	\item $\sigma$ is sufficient to verify that $\Phi$ holds for program $P$;
    \item for each $n \in \mathcal{N}$, the invariant $W\,n$ is true in every state of $\gamma\,(\sigma\,n)$.
\end{enumerate}
In practice, the second condition may not be easy to check since computing $\gamma$ is not always feasible.
Thus, abstract expression evaluation is used instead to perform the validity check, although this is possibly less precise, as the following example shows.
\begin{example}
\label{ex:validity-concrete-vs-abstract}
Assume the non-relational abstract domain $\mathbb{D} = \mathcal{V} \to \mathbb{V}$ of environments, where $\mathbb{V}$ is the abstract domain of individual values.
Using intervals for the latter, let $d = \{\texttt{x} \mapsto [1, 2]\}$ be the computed abstract state at some program location
where $\gamma\,d = \{\{\texttt{x} \mapsto 1\}, \{\texttt{x} \mapsto 2\}\}$ is \emph{exact}.
Consider the validation of the following two logically equivalent invariants at this location:
\begin{enumerate}
    \item The invariant $1 \leq \texttt{x} \land \texttt{x} \leq 2$ holds for each concrete state in $\gamma\,d$.
    It also evaluates abstractly to true on $d$ using standard syntax-driven evaluation (see \cref{sec:propagating-unassume}), because both conjuncts are true for the interval $[1, 2]$.
    \item The invariant $\texttt{x} = 1 \lor \texttt{x} = 2$ also holds for each concrete state in $\gamma\,d$.
    However, when evaluated abstractly and syntax-driven on $d$, it evaluates to an unknown boolean, because both disjuncts evaluate to an unknown boolean for the interval $[1, 2]$.
\end{enumerate}
\end{example}
Hence, the abstract interpreter is not complete, i.e., it may fail to validate witnesses which are indeed valid, due to imprecision arising from abstraction or fixpoint acceleration.
Nevertheless, validating witnesses using abstract expression evaluation is sound, i.e., all witnesses claimed to be validated by the abstract interpreter, are indeed valid.

A witness which maps all locations to the invariant $e_{\textsf{true}}$ is called \emph{trivial}.
The validation of such a witness trivially passes the second validity condition, and checking of the first condition falls entirely to the analyzer itself.
To be useful, a witness has to be non-trivial and aid the analyzer in proving that the program $P$ satisfies the property $\Phi$, either by improving the precision or the performance of the verification process.
For \emph{witness-guided verification}, it suffices if the analyzer can show that $\Phi$ holds ---
even if the invariants of the witness cannot be validated.
Given a witness $(W, P, \Phi)$, the challenge of witness-guided abstract interpretation is
to simultaneously achieve the following:
\begin{enumerate}
	\item	to use the invariants in $W$ to reach a fixpoint $\sigma_W: \mathcal{N} \to \mathbb{D}$ in fewer iterations;
	\item	to avoid overshooting the required property, i.e., $\sigma_W$ suffices to prove $\Phi$;
    \item   to \emph{not trust} the witness, i.e., $\sigma_W$ should remain sound even in presence of \emph{wrong} invariants.
\end{enumerate}
Subsequently, we first consider some intuitive approaches to motivate our unassume operator
for soundly speeding up abstract interpretation with the help of untrusted witnesses.

\section{Initialization-Based Approaches}
\label{sec:initialization-based}
Given a witness $(W, P, \Phi)$, one natural idea is to extract from
the mapping ${W: \mathcal{N} \to \mathcal{E}}$ a mapping $w: \mathcal{N} \to \mathbb{D}$ of initial abstract values as non-$\bot$ start points
for constructing inductive invariants for the program.
We discuss two flavors for realizing this idea, along with their shortcomings.

\subsubsection{Total Initial Values.}
In the first approach, the initial value $w\,n$ for program location $n$ is chosen such that $\gamma\,(w\,n)$ includes every concrete state where $W\,n$ holds.
For example, by choosing $w\,\ell_2 = \{\texttt{x} \mapsto [0, 40]\}$ in \cref{ex:illustrative}.
Such a value, however, is only suitable if \emph{all} relevant information for program location~$n$ is formalized in the invariant $W\,n$
and expressible by the abstract domain.
Apart from trivial cases, both requirements are seldom fulfilled in practice.

For example, consider the invariant $\texttt{*p} \geq 0$ involving a pointer \texttt{p} dereference for some program location $n$.
It provides no information about which variables \texttt{p} may point to, thus nothing can be concluded about any integer variables it intends to describe.
Therefore, $w\,n = \top$ which leads to a complete loss of precision at location $n$ during the analysis.

This approach also makes silent assumptions about the way in which the analyzer computes values,
namely how such initial abstract values are incorporated into analysis, if at all.
For example, TD fixpoint solvers~\cite{Seidl_2021} only use initial values at \emph{dynamically} identified widening points for starting fixpoint iteration.
Additionally, context-sensitive interprocedural analysis is known to give rise to \emph{infinite} constraint systems~\cite{Apinis:2012:SCS},
requiring dedicated changes to the analyzer to ensure that all accessed
constraint variables associated with a given location are appropriately initialized.

\subsubsection{Partial Initial Values.}
In order to remedy the problem that all relevant information must be provided in the invariants for program locations,
one may instead rely on \emph{partial} initialization. For that to work, we assume here that a non-relational abstract domain $\mathbb{D} = \mathcal{V} \to \mathbb{V}$ is used.
Assuming that the invariant~$W\,n$ only speaks of variables from $V \subseteq \mathcal{V}$,
the partial initial value is the same as the total initial value $w\,n$
except all unmentioned variables $x\in\mathcal{V}\setminus V$ are assigned $\bot$.

\begin{example}
\label{ex:unassume-pointer}
Consider for a particular program location $n$, two integer variables \texttt{i} and \texttt{j}
and a pointer variable \texttt{p}.
Let the abstract domain $\mathbb{V}$ of values consist of intervals for abstracting integers and points-to sets for abstracting pointers.
Consider two invariants:
\begin{enumerate}
    \item The witness invariant $\texttt{i} \geq 0 \land \texttt{j} \geq 0$ can be represented by the partial state
	    $\{\texttt{p} \mapsto \bot, \texttt{i} \mapsto [0, \infty], \texttt{j} \mapsto [0, \infty]\}$.
    \item The witness invariant $\texttt{*p} \geq 0$, on the other hand,
	    results in the partial state
	    $\{\texttt{p} \mapsto \top, \texttt{i} \mapsto \bot, \texttt{j} \mapsto \bot\}$.
\end{enumerate}
Now assume that during analysis of the program, the complete abstract state
	$\{\texttt{p} \mapsto \{\texttt{\&i}, \texttt{\&j}\}, \texttt{i} \mapsto [0, 0], \texttt{j} \mapsto [0, 0]\}$, where \texttt{p} may point to either \texttt{i} or \texttt{j},
	reaches the program location $n$.
	In order to exploit the witness, this value is \emph{joined} with the partial state constructed from the witness in the corresponding transfer function.
	For the two invariants above, we respectively obtain:
\begin{enumerate}
    \item $\{\texttt{p} \mapsto \{\texttt{\&i}, \texttt{\&j}\}, \texttt{i} \mapsto [0, \infty], \texttt{j} \mapsto [0, \infty]\}$,
    \item $\{\texttt{p} \mapsto \top, \texttt{i} \mapsto [0, 0], \texttt{j} \mapsto [0, 0]\}$.
\end{enumerate}
The first may be useful to guide the analysis since the information for \texttt{i} and \texttt{j} is maximally relaxed such that the witness invariant
can still be validated, while the information for the pointer variable \texttt{p}
is retained.
On the other hand, the second state loses all information about \texttt{p},
which is problematic if memory is accessed through \texttt{p} later in the program.
At the same time, here, the values for the variables \texttt{i} and \texttt{j} remain overly precise.
Instead, one would have liked to obtain the former abstract state also when using the invariant $\texttt{*p} \geq 0$.
\end{example}

By joining initial values within transfer functions, this approach is more general:
it works for all program locations regardless of the analysis engine and can be seamlessly applied to infinite constraint systems.
Nevertheless, the incorporation of witnesses via partial initial values is only applicable to non-relational domains and cannot depend on analysis state.
Therefore, in the next section, we propose a more general solution that overcomes these issues.

\section{Unassuming}
\label{sec:unassume}

We introduce new statements $\unassume(e)$ to the programming language for all invariants $e \in \mathcal{E}$.
Given a witness $(W, P, \Phi)$, we insert at every location $n$, the statement $\unassume(W\, n)$ if it
is different from $e_\textsf{true}$.
In case the invariant is not a legal program expression, we may instead insert the statement at the location
into the internal representation used by the analyzer (e.g., the control-flow graph).
In the concrete semantics, unassume statements have no effect, i.e., their arguments are not evaluated and thus does not cause runtime errors or undefined behavior.

During the abstract interpretation of the program, the abstract state transformer for the statement
$\unassume(e)$ for location $n$ is meant to inject the desired imprecision into the abstract state for $n$.
Intuitively, the abstract semantics of unassume is dual to the \emph{assume} operation, i.e.,
it relaxes a state instead of refining it. Thus, e.g.,
\[
    \{\texttt{x} \mapsto [0, \infty]\}
    \xrightleftharpoons[\unassume(\texttt{x} \geq 0)]{\assume(\texttt{x} = 0)}
    \{\texttt{x} \mapsto [0, 0]\}.
\]
Note that $\unassume$ is not the inverse of $\assume$ because the used expressions are different.
By integrating unassume operations as statements, they can be treated path- and context-sensitively -- just like all other statements
 -- if the abstract interpreter supports such sensitivity, yielding a general approach.

\subsection{Specification}

Subsequently, we provide
abstract operators $\sems{\unassume_V(e)}: \mathbb{D} \to \mathbb{D}$ which we use to
abstractly interpret the corresponding unassume statement.
The abstract operators are parameterized by the set of variables $V \subseteq \mathcal{V}$
whose values are relaxed up to the constraining invariant $e$.
The abstract unassume operator $\sems{\unassume_V(e)}$
is \emph{sound} if it abstracts the concrete no-op operator,
i.e.,
    \[
	    \gamma\,d \subseteq \gamma\,(\sems{\unassume_V(e)}\,d)
    \]
    for all $d \in \mathbb{D}$.
    In particular, this is the case if the operator is \emph{extensive}, i.e.,
    \[
	    d \sqsubseteq \sems{\unassume_V(e)}\,d.
    \]
Given that the abstract interpreter is sound w.r.t.\ the original program, and sound unassume operations are
inserted, we conclude that the resulting abstract interpreter is sound w.r.t.\ the modified program.
Since the newly inserted statements have no effects in the concrete, the resulting abstract interpreter remains
sound also w.r.t.\ the original program.
This implies the soundness of our validation approach.
\begin{theorem}[Sound witness validation]
    Assume a witness $(W, P, \Phi)$ is used to insert unassume statements and $\sigma_W: \mathcal{N} \to \mathbb{D}$ is the result of analyzing the instrumented program. If the sound analyzer confirms $\Phi$
    and all invariants of $W: \mathcal{N} \to \mathcal{E}$ abstractly evaluate to true in $\sigma_W$,
    then the witness must be valid.
\end{theorem}

\begin{example}
\label{ex:unassume-desired}
	The desired behavior of unassume operators is illustrated by the following examples.
\begin{enumerate}
    \item \label{itm:unassume-desired-unmentioned} Unmentioned parts of the abstract state should be retained:
    \begin{multline*}
        \sems{\unassume_{\{\texttt{x}\}}(\texttt{x} \geq 0)}\,\{\texttt{x} \mapsto [0, 0], \texttt{y} \mapsto [0, 0]\} = \{\texttt{x} \mapsto [0, \infty], \texttt{y} \mapsto [0, 0]\}.
    \end{multline*}
    \item \label{itm:unassume-desired-v} Information on variables used in the invariant, but not
	    contained in $V$ should be retained:
    \begin{multline*}
        \sems{\unassume_{\{\texttt{i}\}}(\texttt{i} \leq \texttt{n})}\,\{\texttt{i} \mapsto [0, 0], \texttt{n} \mapsto [10, 10]\} = \\
        = \{\texttt{i} \mapsto [-\infty, 10], \texttt{n} \mapsto [10, 10]\}.
    \end{multline*}
    \item \label{itm:unassume-desired-rel} Relational invariants between relaxed and not relaxed variables should be preserved whenever possible without restricting the unassumed invariant; e.g.,
    relaxing the state $0 = \texttt{x} \leq \texttt{y}$ with $0 \leq \texttt{x}$ should result in $0 \leq \texttt{x} \leq \texttt{y}$:
    \begin{multline*}
        \sems{\unassume_{\{\texttt{x}\}}(\texttt{x} \geq 0)}\,\{\texttt{x} \leq 0, -\texttt{x} \leq 0, -\texttt{y} \leq 0, \textcolor{gray}{-\texttt{x} - \texttt{y} \leq 0, \texttt{x} - \texttt{y} \leq 0}\} = \\
        = \{-\texttt{x} \leq 0, -\texttt{y} \leq 0, \textcolor{gray}{-\texttt{x} - \texttt{y} \leq 0,{}} \texttt{x} - \texttt{y} \leq 0\}
    \end{multline*}
    when using the octagon domain~\cite{Mine2006}.\footnote{Redundant constraints are grayed out. They can be derived from non-redundant (non-grayed out) constraints using the octagon closure algorithm.}
    More specifically, this is the most precise result which, when projected to $V$, contains the abstract state ${\{-\texttt{x} \leq 0\}}$ defined only by the unassumed invariant.
    \item \label{itm:unassume-desired-pointer} Information provided by the input abstract state should be leveraged to propagate imprecision to further
    variables and heap locations not mentioned in the invariant
    (cf. \cref{ex:unassume-pointer}):
    \begin{multline*}
        \sems{\unassume_{\{\texttt{i}, \texttt{j}\}}(\texttt{*p} \geq 0)}\, \{\texttt{p} \mapsto \{\texttt{\&i}, \texttt{\&j}\}, \texttt{i} \mapsto [0, 0], \texttt{j} \mapsto [0, 0]\} = \\
        = \{\texttt{p} \mapsto \{\texttt{\&i}, \texttt{\&j}\}, \texttt{i} \mapsto [0, \infty], \texttt{j} \mapsto [0, \infty]\}.
    \end{multline*}
\end{enumerate}
We remark that
\cref{itm:unassume-desired-v,itm:unassume-desired-pointer} illustrate cases where $V$ differs from the set of variables
syntactically occurring in $e$.
\end{example}

\subsection{Na\"ive Definition}

We present the first unassume operator in terms of
the abstract operators for non-deterministic assignments and guards.
In this section, we assume the invariant language $\mathcal{E}$ is a subset of the side-effect--free expressions
used for conditional branching in the programming language.

For an expression $e$, let $\assume(e)$ denote the concrete operation which only continues execution if the condition $e$ is true, and aborts otherwise.
Let $\sems{\assume(e)}: \mathbb{D} \to \mathbb{D}$ be a sound abstraction.

For a set of variables $V \subseteq \mathcal{V}$, let $\havoc(V)$ denote the concrete operation which
non-deterministically assigns arbitrary values to all $x\in V$, and
$\sems{\havoc(V)}: \mathbb{D} \to \mathbb{D}$ be a sound abstraction.

\begin{definition}[Na\"ive unassume]
    Let $V \subseteq \mathcal{V}$, $e \in \mathcal{E}$ and $d \in \mathbb{D}$.
	Then the \emph{na\"ive unassume} is defined as
    \[
        \sems{\unassume_V(e)}_1\,d = d \sqcup (\sems{\assume(e)} \circ \sems{\havoc(V)})\,d.
    \]
\end{definition}
Intuitively, the argument state is relaxed by joining with an additional value.
This value is obtained by first forgetting all information about the variables from $V$ and then assuming the
information provided by $e$.
Due to the join, this unassume operator is \emph{sound by construction}.
The na\"ive unassume operator is already sufficient to gain the improvements illustrated by \cref{ex:illustrative} when choosing $V = \{\texttt{x}\}$:
\begin{align*}
    &\phantom{{}={}} \sems{\unassume_{\{\texttt{x}\}}(0 \leq \texttt{x} \leq 40)}_1\,\{\texttt{x} \mapsto [40, 40]\} = \\
    &= \{\texttt{x} \mapsto [40, 40]\} \sqcup (\sems{\assume(0 \leq \texttt{x} \leq 40)} \circ \sems{\havoc(\{\texttt{x}\})})\,\{\texttt{x} \mapsto [40, 40]\} = \\
    &= \{\texttt{x} \mapsto [40, 40]\} \sqcup \sems{\assume(0 \leq \texttt{x} \leq 40)}\,\{\texttt{x} \mapsto \top\} = \\
    &= \{\texttt{x} \mapsto [40, 40]\} \sqcup \{\texttt{x} \mapsto [0, 40]\} = \{\texttt{x} \mapsto [0, 40]\}.
\end{align*}
This operator also
succeeds for \cref{itm:unassume-desired-unmentioned,itm:unassume-desired-v} in \cref{ex:unassume-desired}:
\begin{align*}
    &\phantom{{}={}} \sems{\unassume_{\{\texttt{x}\}}(\texttt{x} \geq 0)}_1\,\{\texttt{x} \mapsto [0, 0], \texttt{y} \mapsto [0, 0]\} = \\
    &= \{\texttt{x} \mapsto [0, 0], \texttt{y} \mapsto [0, 0]\} \sqcup{} \\
    &\qquad {}\sqcup (\sems{\assume(\texttt{x} \geq 0)} \circ \sems{\havoc(\{\texttt{x}\})})\,\{\texttt{x} \mapsto [0, 0], \texttt{y} \mapsto [0, 0]\} = \\
    &= \{\texttt{x} \mapsto [0, 0], \texttt{y} \mapsto [0, 0]\} \sqcup \sems{\assume(\texttt{x} \geq 0)}\,\{\texttt{x} \mapsto \top, \texttt{y} \mapsto [0, 0]\} = \\
    &= \{\texttt{x} \mapsto [0, 0], \texttt{y} \mapsto [0, 0]\} \sqcup \{\texttt{x} \mapsto [0, \infty], \texttt{y} \mapsto [0, 0]\} = \{\texttt{x} \mapsto [0, \infty], \texttt{y} \mapsto [0, 0]\} \\
\intertext{and}
    &\phantom{{}={}} \sems{\unassume_{\{\texttt{i}\}}(\texttt{i} \leq \texttt{n})}_1\,\{\texttt{i} \mapsto [0, 0], \texttt{n} \mapsto [10, 10]\} = \\
    &= \{\texttt{i} \mapsto [0, 0], \texttt{n} \mapsto [10, 10]\} \sqcup{} \\
    &\qquad {}\sqcup (\sems{\assume(\texttt{i} \leq \texttt{n})} \circ \sems{\havoc(\{\texttt{i}\})})\,\{\texttt{i} \mapsto [0, 0], \texttt{n} \mapsto [10, 10]\} = \\
    &= \{\texttt{i} \mapsto [0, 0], \texttt{n} \mapsto [10, 10]\} \sqcup \sems{\assume(\texttt{i} \leq \texttt{n})}\,\{\texttt{i} \mapsto \top, \texttt{n} \mapsto [10, 10]\} = \\
    &= \{\texttt{i} \mapsto [0, 0], \texttt{n} \mapsto [10, 10]\} \sqcup \{\texttt{i} \mapsto [-\infty, 10], \texttt{n} \mapsto [10, 10]\} = \\
    &= \{\texttt{i} \mapsto [-\infty, 10], \texttt{n} \mapsto [10, 10]\}.
\end{align*}
But it fails when there are relations between
elements of $V$ and $\mathcal{V}\setminus V$, e.g., for \cref{itm:unassume-desired-rel} with $d = \{\texttt{x} \leq 0, -\texttt{x} \leq 0, -\texttt{y} \leq 0, \textcolor{gray}{-\texttt{x} - \texttt{y} \leq 0, \texttt{x} - \texttt{y} \leq 0}\}$:
\begin{align*}
    &\phantom{{}={}} \sems{\unassume_{\{\texttt{x}\}}(\texttt{x} \geq 0)}_1\,d = d \sqcup (\sems{\assume(\texttt{x} \geq 0)} \circ \sems{\havoc(\{\texttt{x}\})})\,d = \\
    &= d \sqcup \sems{\assume(\texttt{x} \geq 0)}\,\{-\texttt{y} \leq 0\} = d \sqcup \{-\texttt{x} \leq 0, -\texttt{y} \leq 0, \textcolor{gray}{-\texttt{x} - \texttt{y} \leq 0}\} = \\
    &= \{-\texttt{x} \leq 0, -\texttt{y} \leq 0, \textcolor{gray}{-\texttt{x} - \texttt{y} \leq 0}\}.
\end{align*}
In this case the octagon constraint $\texttt{x} - \texttt{y} \leq 0$ is lost by $\havoc$ing and cannot be recovered by assuming.

\subsection{Dual-Narrowing}
\label{sec:dualnarrowing-unassume}
We will address the above challenge by relying on additional insights from abstract interpretation.
Let us recall the term dual-narrowing, which is the lattice analogue of Craig interpolation~\cite{10.1007/978-3-662-46081-8_2}.
A dual-narrowing operator $\dualnarrow: \mathbb{D} \rightarrow \mathbb{D} \rightarrow \mathbb{D}$ returns for every $d_1, d_2 \in \mathbb{D}$ with
$d_1 \sqleq d_2$,  a value between both of them, i.e., $d_1 \sqleq d_1 \dualnarrow d_2 \sqleq d_2$.

Using such an operator, we can define an abstract unassume that, given $d$, may return an abstract value in the range from $d$ to $\sems{\unassume_V(e)}_1\,d$:
\begin{definition}[Dual-narrowing unassume]
    Let $\dualnarrow: \mathbb{D} \rightarrow \mathbb{D} \rightarrow \mathbb{D}$ be a dual-narrowing. Let $V \subseteq \mathcal{V}$, $e \in \mathcal{E}$ and $d \in \mathbb{D}$.
    Then the \emph{dual-narrowing unassume} is defined as a wrapper around the na\"ive unassume:
    \[
        \sems{\unassume_V(e)}_2\,d = d \dualnarrow \sems{\unassume_V(e)}_1\,d.
    \]
\end{definition}
\begin{example}
\label{ex:unassume-standard-rel-fix}
A dual-narrowing for relational domains can be defined using \emph{heterogeneous} environments and \emph{strengthening}~\cite{Journault_2019}.
Let $\dom(d) \subseteq \mathcal{V}$ denote the environment of the abstract value $d \in \mathbb{D}$.
Let $d\restrict{V}$ denote the restriction of the abstract value $d \in \mathbb{D}$ to the program variables $V \subseteq \mathcal{V}$.

An environment-aware order $\sqleqstar$ is defined for $d_1, d_2 \in \mathbb{D}$ by
\[
    d_1 \sqleqstar d_2 \quad\Longleftrightarrow\quad \dom(d_1) \subseteq \dom(d_2) \land d_1 \sqleq d_2\restrict{\dom(d_1)}.
\]
Let $\sqcupstar: \mathbb{D} \to \mathbb{D} \to \mathbb{D}$ be an upper bound
operator w.r.t. $\sqleqstar$, such that the resulting environment is minimal, i.e., $\dom(d_1 \sqcupstar d_2) = \dom(d_1) \cup \dom(d_2)$.
Specifically, \citet{Journault_2019} define $\sqcupstar$ as follows.
The result of joining $d_1$ and $d_2$ in their common environment $\dom(d_1) \cap \dom(d_2)$ is extended to $\dom(d_1) \cup \dom(d_2)$ by adding unconstrained dimensions.
A \emph{strengthening} operator refines this result by iteratively adding back constraints from both arguments which would not cause the upper-boundedness w.r.t.\ $\sqleqstar$ to be violated.
Note that this definition is not semantic, i.e., the result depends on the constraints representing the arguments and their processing order.

By defining $d_1 \dualnarrow d_2 = d_1 \sqcupstar d_2\restrict{V}$, which is \emph{parametrized} by $V$,
dual-narrowing unassume yields the following desired result for \cref{itm:unassume-desired-rel} from \cref{ex:unassume-desired} with $d = \{\texttt{x} \leq 0, -\texttt{x} \leq 0, -\texttt{y} \leq 0, \textcolor{gray}{-\texttt{x} - \texttt{y} \leq 0, \texttt{x} - \texttt{y} \leq 0}\}$:
\begin{align*}
    &\phantom{{}={}} \sems{\unassume_{\{\texttt{x}\}}(\texttt{x} \geq 0)}_2\,d = d \dualnarrow \sems{\unassume_{\{\texttt{x}\}}(\texttt{x} \geq 0)}_1\,d = \\
    &= d \dualnarrow \{-\texttt{x} \leq 0, -\texttt{y} \leq 0, \textcolor{gray}{-\texttt{x} - \texttt{y} \leq 0}\} = d \sqcupstar \{-\texttt{x} \leq 0, -\texttt{y} \leq 0, \textcolor{gray}{-\texttt{x} - \texttt{y} \leq 0}\}\restrict{\{\texttt{x}\}} = \\
    &= d \sqcupstar \{-\texttt{x} \leq 0\} = \{-\texttt{x} \leq 0, -\texttt{y} \leq 0, \textcolor{gray}{-\texttt{x} - \texttt{y} \leq 0,{}} \texttt{x} - \texttt{y} \leq 0\}.
\end{align*}
Although the restriction to $V$ first destroys relations between $V$ and $\mathcal{V} \setminus V$, the subsequent strengthening join can restore original relations which are compatible with $e$ on $V$.
\end{example}

\section{Unassuming Indirectly}
\label{sec:indirect-unassume}

\noindent
We now turn to the unassuming of more complex invariants, which include indirection via pointers and dependent subexpressions.
Na\"ive unassume is unable to achieve the desired precision for \cref{itm:unassume-desired-pointer} from \cref{ex:unassume-desired} with $d = \{\texttt{p} \mapsto \{\texttt{\&i}, \texttt{\&j}\}, \texttt{i} \mapsto [0, 0], \texttt{j} \mapsto [0, 0]\}$:
\begin{align*}
    &\phantom{{}={}} \sems{\unassume_{\{\texttt{i}, \texttt{j}\}}(\texttt{*p} \geq 0)}_1\, \{\texttt{p} \mapsto \{\texttt{\&i}, \texttt{\&j}\}, d = \\
    &= d \sqcup (\sems{\assume(\texttt{*p} \geq 0)} \circ \sems{\havoc(\{\texttt{i}, \texttt{j}\})})\,d = \\
    &= d \sqcup \sems{\assume(\texttt{*p} \geq 0)}\,\{\texttt{p} \mapsto \{\texttt{\&i}, \texttt{\&j}\}, \texttt{i} \mapsto \top, \texttt{j} \mapsto \top\} = \\
    &= d \sqcup \{\texttt{p} \mapsto \{\texttt{\&i}, \texttt{\&j}\}, \texttt{i} \mapsto \top, \texttt{j} \mapsto \top\} = \{\texttt{p} \mapsto \{\texttt{\&i}, \texttt{\&j}\}, \texttt{i} \mapsto \top, \texttt{j} \mapsto \top\}.
\end{align*}
This is due to both integer variables being havoced and the assume operator not being able to soundly refine via ambiguous may-point-to sets (see \cref{app:pointers}).
Technically, there exists a dual-narrowing that yields the desired result, but it would be ad-hoc.

To address the disjunctive nature of the may-point-to set, we propose an improved unassume operator.
Suppose we
are provided a family of mappings ${\explode_V(e) : \mathbb{D} \to 2^\mathbb{D}}$
which explode any given abstract state $d$ into a non-empty finite subset
${\explode_V(e)\,d \subseteq \mathbb{D}}$ of abstract states where for each resulting element $d'$
we have $d' \sqsubseteq d$.
The \emph{explode} operator can be used to make disjunctive information in abstract states explicit,
e.g., resolve non-singleton may-point-to sets for pointer variables not contained in $V$.
\begin{definition}[Exploding unassume]
    Let $V \subseteq \mathcal{V}$, $e \in \mathcal{E}$ and $d \in \mathbb{D}$.
    Let $\explode_V(e)$ be an explode operator.
    Then the \emph{exploding unassume} is defined as
    \[
        \sems{\unassume_V(e)}_3\,d =
        \bigsqcap_{\mathclap{d' \in \explode_V(e)\, d}} d \sqcup
        (\sems{\assume(e)} \circ \sems{\havoc(V)})\,d'.
    \]
\end{definition}
This improved unassume operator is extensive and therefore sound for any choice of $\explode_V(e)$.
One might want to establish that ${\bigsqcup \explode_V(e)\,d = d}$ holds, but this is not necessary for soundness.
Whereas $\explode_V(e)\,d = \{\bot\}$ would make the unassume a no-op.

\begin{example}
\label{ex:extract}
Consider the following explode operator, which splits ambiguous may-point-to sets:
\begin{multline*}
    \explode_{\{\texttt{i}, \texttt{j}\}}(\texttt{*p} \geq 0)\, \{\texttt{p} \mapsto \{\texttt{\&i}, \texttt{\&j}\}, \texttt{i} \mapsto [0, 0], \texttt{j} \mapsto [0, 0]\} = \\
    = \{\{\texttt{p} \mapsto \{\texttt{\&i}\}, \texttt{i} \mapsto [0, 0], \texttt{j} \mapsto [0, 0]\}, \{\texttt{p} \mapsto \{\texttt{\&j}\}, \texttt{i} \mapsto [0, 0], \texttt{j} \mapsto [0, 0]\}\}.
\end{multline*}
Using this explode operator, \cref{itm:unassume-desired-pointer} from \cref{ex:unassume-desired} is handled as desired with $d = \{\texttt{p} \mapsto \{\texttt{\&i}, \texttt{\&j}\}, \texttt{i} \mapsto [0, 0], \texttt{j} \mapsto [0, 0]\}$:
\begin{align*}
    &\phantom{{}={}} \sems{\unassume_{\{\texttt{i}, \texttt{j}\}}(\texttt{*p} \geq 0)}_3\, d = \\
    &= (d \sqcup (\sems{\assume(\texttt{*p} \geq 0)} \circ \sems{\havoc(\{\texttt{i}, \texttt{j}\})})\,\{\texttt{p} \mapsto \{\texttt{\&i}\}, \texttt{i} \mapsto [0, 0], \texttt{j} \mapsto [0, 0]\}) \sqcap{} \\
    &\quad {}\sqcap (d \sqcup (\sems{\assume(\texttt{*p} \geq 0)} \circ \sems{\havoc(\{\texttt{i}, \texttt{j}\})})\,\{\texttt{p} \mapsto \{\texttt{\&j}\}, \texttt{i} \mapsto [0, 0], \texttt{j} \mapsto [0, 0]\}) = \\
    &= (d \sqcup \sems{\assume(\texttt{*p} \geq 0)}\,\{\texttt{p} \mapsto \{\texttt{\&i}\}, \texttt{i} \mapsto \top, \texttt{j} \mapsto \top\}) \sqcap{} \\
    &\quad {}\sqcap (d \sqcup \sems{\assume(\texttt{*p} \geq 0)}\,\{\texttt{p} \mapsto \{\texttt{\&j}\}, \texttt{i} \mapsto \top, \texttt{j} \mapsto \top\}) = \\
    &= (d \sqcup \{\texttt{p} \mapsto \{\texttt{\&i}\}, \texttt{i} \mapsto [0, \infty], \texttt{j} \mapsto \top\}) \sqcap{} \\
    &\quad {}\sqcap (d \sqcup \{\texttt{p} \mapsto \{\texttt{\&j}\}, \texttt{i} \mapsto \top, \texttt{j} \mapsto [0, \infty]\}) = \\
    &= \{\texttt{p} \mapsto \{\texttt{\&i}, \texttt{\&j}\}, \texttt{i} \mapsto [0, \infty], \texttt{j} \mapsto \top\} \sqcap \{\texttt{p} \mapsto \{\texttt{\&i}, \texttt{\&j}\}, \texttt{i} \mapsto \top, \texttt{j} \mapsto [0, \infty]\} = \\
    &= \{\texttt{p} \mapsto \{\texttt{\&i}, \texttt{\&j}\}, \texttt{i} \mapsto [0, \infty], \texttt{j} \mapsto [0, \infty]\}.
\end{align*}
\end{example}
\begin{example}
\label{ex:exploding-unassume-dependent}
However, consider the following, where different subexpressions depend on each other (here through \texttt{p}):
\[
    \sems{\unassume_{\{\texttt{p}, \texttt{i}, \texttt{j}\}}((\texttt{p} = \texttt{\&i} \lor \texttt{p} = \texttt{\&j}) \land \texttt{*p} \geq 0)}\, \{\texttt{p} \mapsto \{\texttt{\&i}\}, \texttt{i} \mapsto [0, 0], \texttt{j} \mapsto [0, 0]\}.
\]
In contrast to \cref{ex:extract}, there is no ambiguous may-point-to set in the abstract state supplied as the argument.
All possible explosions lead to the same issue as when using the na\"ive unassume on this example.
After havocing, the environment contains $\texttt{p} \mapsto \top$, thus, in the assume a top pointer needs to be dereferenced and its targets refined.
The semantics of this is unclear and imprecise at best, when one has to consider assignments to \emph{all} possible (unrelated) memory locations.
\end{example}

\subsection{Propagating Unassume}
\label{sec:propagating-unassume}

The \emph{HC4-revise} algorithm by \citet{Benhamou99} can be used to implement the assume operation for complex expressions on non-relational domains in a syntax-directed manner~\cite{Mine2017,ziat:tel-03987752}.
It is also known as \emph{backwards evaluation}~\cite{Cousot98-5}.
We describe the algorithm and then apply it to construct an unassume operator.

We loosely follow the presentation by \citet{Cousot98-5}.
Let the languages of expressions $e$ and logical conditions $c$ be defined by the grammars in \cref{fig:exp-grammar}.
For each $n \in \mathbb{N}$, let $\mathcal{O}_n$ be the set of $n$-ary operators.
For simplicity of presentation, assume that the condition is in negation normal form (NNF), i.e., negations in conditions have been ``pushed down'' into binary comparisons according to boolean logic.
The logical conditions form an invariant language (see \cref{sec:preliminaries}).
The following algorithms generalize from just variables to lvalues, allowing for languages with pointers like our example invariant language from before.
This generalization is formalized in \cref{app:pointers}.

\begin{figure*}
    \centering
    \begin{subfigure}[t]{0.45\textwidth}
        \begin{align*}
            e ::=\; & k & \text{(constant)} \\
            |\; & x & \text{(variable, $x \in \mathcal{V}$)} \\
            |\; & \square\,(e)_{i = 1}^n & \begin{aligned}[t]
                    & (\text{$n$-ary operator}, \\
                    & \quad n \in \mathbb{N}, \square \in \mathcal{O}_n)
                \end{aligned}
        \end{align*}
    \end{subfigure}
    ~
    \begin{subfigure}[t]{0.5\textwidth}
        \begin{align*}
            c ::=\; & e \bowtie e & \begin{aligned}[t]
                    & (\text{binary comparison}, \\
                    & \quad {\bowtie} \in \{{=}, {\neq}, {<}, {\leq}, {>}, {\geq}\})
                \end{aligned} \\
            |\; & c \land c & \text{(conjunction)} \\
            |\; & c \lor c & \text{(disjunction)}
        \end{align*}
    \end{subfigure}
    \caption{Syntax of expressions and conditions.}
    \label{fig:exp-grammar}
\end{figure*}

\begin{figure*}
    \centering
    \begin{subfigure}{0.4\textwidth}
        \begin{align*}
            &\sems[\mathbb{E}]{e} : \mathbb{D} \to \mathbb{V} \\
            &\sems[\mathbb{E}]{k}\,d = k^\sharp \\
            &\sems[\mathbb{E}]{x}\,d = d\,x \\
            &\sems[\mathbb{E}]{\square\,(e_i)_{i = 1}^n}\,d = \square^\sharp\,(\sems[\mathbb{E}]{e_i}\,d)_{i = 1}^n
        \end{align*}
    \end{subfigure}
    ~
    \begin{subfigure}{0.5\textwidth}
        \begin{align*}
            &\sems[\mathbb{C}]{c} : \mathbb{D} \to \mathbb{B} \\
            &\sems[\mathbb{C}]{e_1 \bowtie e_2}\,d = \sems[\mathbb{E}]{e_1}\,d \bowtie^\sharp \sems[\mathbb{E}]{e_2}\,d \\
            &\sems[\mathbb{C}]{c_1 \land c_2}\,d = \sems[\mathbb{C}]{c_1}\,d \land^\sharp \sems[\mathbb{C}]{c_2}\,d \\
            &\sems[\mathbb{C}]{c_1 \lor c_2}\,d = \sems[\mathbb{C}]{c_1}\,d \lor^\sharp \sems[\mathbb{C}]{c_2}\,d
        \end{align*}
    \end{subfigure}
    \caption{Forward evaluation of expressions and conditions.}
    \label{fig:exp-eval}
\end{figure*}

\paragraph{Evaluation.}
Let $\mathbb{V}$ be the abstract domain for individual values and ${\mathbb{D} = \mathcal{V} \to \mathbb{V}}$ the abstract domain for non-relational environments.
Let $\mathbb{B}$ be the flat boolean domain, where $\bot \sqleq \{\textsf{true}^\sharp, \textsf{false}^\sharp\} \sqleq \top$.
The standard abstract forward evaluation of expressions $\sems[\mathbb{E}]{e}$ and conditions $\sems[\mathbb{C}]{c}$ in the non-relational environment $d \in \mathbb{D}$ is shown in \cref{fig:exp-eval}.
For a constant $k$, let $k^\sharp$ be its corresponding abstraction, and $\square^\sharp$, $\bowtie^\sharp$, $\land^\sharp$, $\lor^\sharp$ be abstract versions of the corresponding operators.

\vspace{10pt} % TODO: nudge line to next page
\subsubsection{Assume.}
The HC4-revise algorithm for the assume operation has two phases:
\begin{enumerate}
    \item Bottom-up forward propagation on the expression tree abstractly evaluates the expression, as usual.
    \item Top-down backward propagation refines each abstract value with the expected result of the sub-expression.
    This relies on backward abstract operators, which refine each argument based on the other arguments and the expected result, while variables are refined at the leaves.
\end{enumerate}
The algorithm $\sems{\assume(e)}$ with its abstract backward evaluation of expressions $\sems[\overleftarrow{\mathbb{E}}]{e}$ and conditions $\sems[\overleftarrow{\mathbb{C}}]{c}$ is shown in \cref{fig:exp-assume}.
Instead of evaluating to an abstract value, they refine values of variables in the abstract environment.
For each $n \in \mathbb{N}$, $\square \in \mathcal{O}_n$, let $\overleftarrow{\square}^\sharp: \mathbb{V} \to \mathbb{V}^n \to \mathbb{V}^n$ be the abstract backward version of the $n$-ary operator $\square$.
It returns abstract values for its arguments under the assumption that the operator evaluates to the given abstract value $v'$ and the other arguments have the given abstract values.
For example, if $n = 2$, then
\begin{align*}
    \overleftarrow{\square}^\sharp\,v'\,(v_1, v_2) = (v_1', v_2') \implies {} &\{x_1 \in \gamma\,\mathbb{V} \mid \exists x_2 \in \gamma\,v_2: \square\,(x_1, x_2) \in \gamma\,v'\} \subseteq \gamma\,v_1' \land {} \\
    {} \land {} &\{x_2 \in \gamma\,\mathbb{V} \mid \exists x_1 \in \gamma\,v_1: \square\,(x_1, x_2) \in \gamma\,v'\} \subseteq \gamma\,v_2'.
\end{align*}
Unlike \citet{Cousot98-5} and \citet{Mine2017}, we require that the backward operators \emph{do not} intersect an argument's backward-computed value with its current value.
Instead, we make this explicit in the algorithm like \citet{Benhamou99}.
Similarly, let $\overleftarrow{\bowtie}^\sharp: \mathbb{V} \to \mathbb{V} \to \mathbb{V} \times \mathbb{V}$ be the abstract backward version of the comparison $\bowtie$.
Since conditions are in NNF, the expected result is always $\textsf{true}^\sharp$ and no $v'$ argument is needed for it.
The evaluations $\sems[\mathbb{E}]{e}\,d$ should all be cached and reused from a single forward evaluation as the argument environment is passed around without changes~\cite{Benhamou99,Mine2017}.

\begin{comment}
\begin{figure*}
    \begin{minted}[escapeinside=//]{ocaml}
        val eval: expr -> state -> value

        let rec assume' (eval0: expr -> value) (e: expr) (v': value) (s: state): state =
          match e with
          | Const _ -> s
          | Var x -> s[x -> s[x] /$\sqcap$/ v']
          | Unary (/$\square$/, e1) ->
            let v1 = eval0 e1 in
            let v'1 = /$\overleftarrow{\square}$/ v1 v' in
            assume' eval0 e1 (v1 /$\sqcap$/ v'1) s
          | Binary (/$\square$/, e1, e2) ->
            let (v1, v2) = (eval0 e1, eval0 e2) in
            let (v'1, v'2) = /$\overleftarrow{\square}$/ v1 v2 v' in
            assume' eval0 e2 (v2 /$\sqcap$/ v'2) (assume' eval0 e1 (v1 /$\sqcap$/ v'1) s)

        let assume (e: expr) (s: state): state =
          let eval0 e = eval e s in (* memoized *)
          assume' eval0 e true s
    \end{minted}
    \caption{Propagation assume algorithm}
\end{figure*}
\end{comment}

\begin{figure*}
    \[
        \sems{\assume(e)}\,d = \sems[\overleftarrow{\mathbb{C}}]{e}\,d
    \]
    where
    \begin{center}
    \vspace*{-2em}
    \begin{subfigure}[t]{0.4\textwidth}
        \begin{align*}
            &\sems[\overleftarrow{\mathbb{E}}]{e} : \mathbb{V} \to \mathbb{D} \to \mathbb{D} \\
            &\sems[\overleftarrow{\mathbb{E}}]{k}\,v'\,d = \mathbf{if}\;k^\sharp \sqleq v'\;\mathbf{then}\;d\;\mathbf{else}\;\bot \\
            &\sems[\overleftarrow{\mathbb{E}}]{x}\,v'\,d = d[x \mapsto d\,x \sqcap v'] \\
            &\sems[\overleftarrow{\mathbb{E}}]{\square\,(e_i)_{i = 1}^n}\,v'\,d = \\
            &\qquad \begin{aligned}[t]
                    & \mathbf{let}\;(v_i)_{i = 1}^n = (\sems[\mathbb{E}]{e_i}\,d)_{i = 1}^n\;\mathbf{in} \\
                    & \mathbf{let}\;(v'_i)_{i = 1}^n = \overleftarrow{\square}^\sharp\,v'\,(v_i)_{i = 1}^n\;\mathbf{in} \\
                    & \bigsqcap_{i = 1}^n \sems[\overleftarrow{\mathbb{E}}]{e_i}\,(v_i \sqcap v'_i)\,d
                \end{aligned}
        \end{align*}
    \end{subfigure}
    ~
    \begin{subfigure}[t]{0.5\textwidth}
        \begin{align*}
            &\sems[\overleftarrow{\mathbb{C}}]{c} : \mathbb{D} \to \mathbb{D} \\
            &\sems[\overleftarrow{\mathbb{C}}]{e_1 \bowtie e_2}\,d = \\
            &\qquad \begin{aligned}[t]
                    & \mathbf{let}\;(v_1, v_2) = (\sems[\mathbb{E}]{e_1}\,d, \sems[\mathbb{E}]{e_2}\,d)\;\mathbf{in} \\
                    & \mathbf{let}\;(v'_1, v'_2) = v_1 \mathrel{\overleftarrow{\bowtie}^\sharp} v_2\;\mathbf{in} \\
                    & \sems[\overleftarrow{\mathbb{E}}]{e_1}\,(v_1 \sqcap v'_1)\,d \sqcap \sems[\overleftarrow{\mathbb{E}}]{e_2}\,(v_2 \sqcap v'_2)\,d
                \end{aligned} \\
            &\sems[\overleftarrow{\mathbb{C}}]{c_1 \land c_2}\,d = \sems[\overleftarrow{\mathbb{C}}]{c_1}\,d \sqcap \sems[\overleftarrow{\mathbb{C}}]{c_2}\,d \\
            &\sems[\overleftarrow{\mathbb{C}}]{c_1 \lor c_2}\,d = \sems[\overleftarrow{\mathbb{C}}]{c_1}\,d \sqcup \sems[\overleftarrow{\mathbb{C}}]{c_2}\,d
        \end{align*}
    \end{subfigure}
    \end{center}
    \vspace*{-1em}
    \caption{Assume via backward evaluation of expressions and conditions by the propagation algorithm.}
    \label{fig:exp-assume}
\end{figure*}

\subsubsection{Unassume.}
This algorithm can be adapted into a \emph{propagating unassume} operator $\sems{\unassume(e)}$ as shown in \cref{fig:exp-unassume}.
Changes are required to achieve the following properties:
\begin{description}
    \item[Variable set variance.]
    In \cref{ex:exploding-unassume-dependent} the first conjunct should relax $\{\texttt{p}\}$, while the second should relax $\{\texttt{i}, \texttt{j}\}$ via the relaxed pointer.
    In order to allow different sub-expressions to relax different variable sets,
    the abstract environments returned by $\sems[\tilde{\mathbb{E}}]{e}$ and $\sems[\tilde{\mathbb{C}}]{c}$ are partial: they only contain variables which have been relaxed at leaves in the corresponding sub-expression.

    Thus heterogeneous lattice join $\sqcupstar$ from \cref{ex:unassume-standard-rel-fix} is used.
    However, here in the non-relational case its definition is simpler:
    values are joined pointwise while using $\bot$ for missing variables.
    The heterogeneous lattice meet $\sqcapbullet$ is defined analogously, using $\top$ for missing variables.
    Note that $\sqcapbullet$ is \emph{not} the meet w.r.t.\ $\sqleqstar$, because $\sqcapbullet$ must preserve all relaxed variables from both operands, not just the common ones.

    \item[Soundness.]
    The result is joined with the pre-unassume environment to ensure soundness.

    \item[Relaxation.]
    Backward propagation only propagates backward values $v_i'$ and does not refine them using abstract values $v_i$ computed by forward propagation.
    Otherwise, sub-expressions cannot be relaxed at all from their current values.
    Note that the forward values are still necessary for evaluating the backward operators.
    This modification on its own yields the HC4-revise${}^\star$ algorithm described by \citet{Benhamou99}.
\end{description}

\definecolor{lightyellow}{cmyk}{0,0,0.3,0}
% https://tex.stackexchange.com/a/33402
\newcommand{\changed}[1]{\mathchoice%
    {\colorbox{lightyellow}{$\displaystyle#1$}}%
    {\colorbox{lightyellow}{$\textstyle#1$}}%
    {\colorbox{lightyellow}{$\scriptstyle#1$}}%
    {\colorbox{lightyellow}{$\scriptscriptstyle#1$}}}
\begin{figure*}
    \[
        \sems{\unassume(e)}\,d = \changed{d \sqcupstar {}} (\sems[\tilde{\mathbb{C}}]{e}\,d)
    \]
    where
    \begin{center}
    \vspace*{-2em}
    \begin{subfigure}[t]{0.4\textwidth}
        \begin{align*}
            &\sems[\tilde{\mathbb{E}}]{e} : \mathbb{V} \to \mathbb{D} \to \mathbb{D} \\
            &\sems[\tilde{\mathbb{E}}]{k}\,v'\,d = \changed{\emptyset} \\
            &\sems[\tilde{\mathbb{E}}]{x}\,v'\,d = \changed{\{x \mapsto v'\}} \\
            &\sems[\tilde{\mathbb{E}}]{\square\,(e_i)_{i = 1}^n}\,v'\,d = \\
            &\qquad \begin{aligned}[t]
                    & \mathbf{let}\;(v_i)_{i = 1}^n = (\sems[\mathbb{E}]{e_i}\,d)_{i = 1}^n\;\mathbf{in} \\
                    & \mathbf{let}\;(v'_i)_{i = 1}^n = \overleftarrow{\square}^\sharp\,v'\,(v_i)_{i = 1}^n\;\mathbf{in} \\
                    & \mathop{\changed{\bigsqcapbullet}}\limits_{i = 1}^n \sems[\tilde{\mathbb{E}}]{e_i}\,\changed{v'_i}\,d
                \end{aligned}
        \end{align*}
    \end{subfigure}
    ~
    \begin{subfigure}[t]{0.5\textwidth}
        \begin{align*}
            &\sems[\tilde{\mathbb{C}}]{c} : \mathbb{D} \to \mathbb{D} \\
            &\sems[\tilde{\mathbb{C}}]{e_1 \bowtie e_2}\,d = \\
            &\qquad \begin{aligned}[t]
                    & \mathbf{let}\;(v_1, v_2) = (\sems[\mathbb{E}]{e_1}\,d, \sems[\mathbb{E}]{e_2}\,d)\;\mathbf{in} \\
                    & \mathbf{let}\;(v'_1, v'_2) = v_1 \mathrel{\overleftarrow{\bowtie}^\sharp} v_2\;\mathbf{in} \\
                    & \sems[\tilde{\mathbb{E}}]{e_1}\,\changed{v'_1}\,d \changed{{} \sqcapbullet {}} \sems[\tilde{\mathbb{E}}]{e_2}\,\changed{v'_2}\,d
                \end{aligned} \\
            &\sems[\tilde{\mathbb{C}}]{c_1 \land c_2}\,d = \sems[\tilde{\mathbb{C}}]{c_1}\,d \changed{{} \sqcapbullet {}} \sems[\tilde{\mathbb{C}}]{c_2}\,d \\
            &\sems[\tilde{\mathbb{C}}]{c_1 \lor c_2}\,d = \sems[\tilde{\mathbb{C}}]{c_1}\,d \changed{{} \sqcupstar {}} \sems[\tilde{\mathbb{C}}]{c_2}\,d
        \end{align*}
    \end{subfigure}
    \end{center}
    \vspace*{-1em}
    \caption{Unassume via backward evaluation of expressions and conditions by the propagation algorithm (changes from the assume algorithm are $\changed{\text{highlighted}}$).}
    \label{fig:exp-unassume}
\end{figure*}

\noindent
Unlike our previous unassume operators, this algorithm implicitly chooses $V$ to be those variables which are relaxed in the process.
Note that it is different from the set of variables syntactically occurring in $e$ in a more complex invariant language, such as our example language with pointers, which is described in \cref{app:pointers}.
This is illustrated by \cref{ex:propagating-unassume-dependent} below.

\paragraph{Local Iteration.}
Repeated application of the propagation algorithm for assuming can improve precision in the presence of dependent subexpressions, i.e., when the same variable occurs multiple times in the condition~\cite{Cousot98-5,Mine2017}.
Analogously, repeated application of the propagation algorithm for unassuming can perform more relaxation in the presence of dependent subexpressions.
Both repetitions can be iterated to a local fixpoint.

\begin{example}
\label{ex:propagating-unassume-dependent}
Consider using the above algorithm for the case from \cref{ex:exploding-unassume-dependent}:
\[
    \sems{\unassume((\texttt{p} = \texttt{\&i} \lor \texttt{p} = \texttt{\&j}) \land \texttt{*p} \geq 0)}\, \{\texttt{p} \mapsto \{\texttt{\&i}\}, \texttt{i} \mapsto [0, 0], \texttt{j} \mapsto [0, 0]\}.
\]
As formalized in \cref{app:pointers}, the first backward propagation returns for ${\texttt{p} = \texttt{\&i}} \lor \texttt{p} = \texttt{\&j}$ the partial map $\{\texttt{p} \mapsto \{\texttt{\&i}, \texttt{\&j}\}\}$ and uses $\texttt{*p} \geq 0$ to relax \texttt{*p}.
To do so, backward operator $\overleftarrow{\geq}^\sharp$ uses the expected true result and its forward-evaluated right argument $[0, 0]$ to propagate the expected value $[0, \infty]$ into its left argument.
Using forward evaluated $\texttt{p} \mapsto \{\texttt{\&i}\}$, backward propagation of the lvalue \texttt{*p}, acting as a leaf, returns the partial map $\{\texttt{i} \mapsto [0, \infty]\}$.
The new constructed environment is $\{\texttt{p} \mapsto \{\texttt{\&i}, \texttt{\&j}\}, \texttt{i} \mapsto [0, \infty], \texttt{j} \mapsto [0, 0]\}$.

The second backward propagation does all of the above and also includes $\texttt{j} \mapsto [0, \infty]$, due to the new points-to set.
The final fixpoint environment is $\{\texttt{p} \mapsto \{\texttt{\&i}, \texttt{\&j}\}, \texttt{i} \mapsto [0, \infty], \texttt{j} \mapsto [0, \infty]\}$.
\end{example}
In \cref{ex:propagating-unassume-dependent} the two iterations induced $V = \{\texttt{p}, \texttt{i}, \texttt{j}\}$, which used with na\"ive unassume yields the issue described in \cref{ex:exploding-unassume-dependent}.
Therefore propagating unassume is not equivalent to simply using its induced variable set with a na\"ive unassume.
Propagating unassume fuses the multiple steps involved together into one algorithm which avoids intermediate imprecision and undefined behavior.
We do not give an exact characterization of the result computed by the modified algorithm as it has remained an open problem for HC4-revise itself~\cite{Benhamou99,Goualard2005}.

In case the value lattice has infinite chains, the local iteration of propagating assume must use narrowing to ensure termination~\cite{Cousot98-5,Mine2017}.
Similarly, the local iteration of propagating unassume must instead use widening.
\begin{example}
Consider using the above algorithm to compute the following:
\[
    \sems{\unassume(\texttt{i} \leq \texttt{i} + 1)}\,\{\texttt{i} \mapsto [0, 0]\}.
\]
The algorithm makes following iterations:
\begin{enumerate}
    \item The first forward propagation computes $[0, 0] \leq^\sharp [1, 1]$.
    Backward propagation then returns $\{\texttt{i} \mapsto [-\infty, 1] \sqcap [-1, \infty] = [-1, 1]\}$.

    \item The second forward propagation computes $[-1, 1] \leq^\sharp [0, 2]$.
    Backward propagation then returns $\{\texttt{i} \mapsto [-\infty, 2] \sqcap [-2, \infty] = [-2, 2]\}$.
    If no widening is applied, then these bounds keep growing by one per iteration.
    If widening is applied, then we get $[-1, 1] \widen [-2, 2] = \top$.

    \item The third forward propagation computes $\top \leq^\sharp \top$.
    Backward propagation cannot relax anything further, so the result is $\{\texttt{i} \mapsto \top\}$.
    This is consistent with expectation: all values of \texttt{i}, where the tautology holds.
\end{enumerate}
\end{example}

\section{Evaluation}
\label{sec:evaluation}

We implement the unassume operator in a state-of-the-art abstract interpreter.
Since the analyzer is sound, this yields a sound witness validator.
However, a sound validator can trivially be obtained by replacing the unassume operator with the identity operator and ignoring the witnesses entirely.
Therefore, our experimental evaluation aims to demonstrate that our witness-guided verifier effectively uses witnesses.
More specifically, we seek to confirm that
\enquote{the effort and feasibility of validation depends on witness content}~\cite{Beyer2016}.
To assess the analyzer's dependency on witness content, we pose the following questions:
\begin{description}
    \item[Precision.] Can the witness-guided verifier leverage witnesses to validate verification results that it could not confirm without a witness?
    \item[Performance.] Do witnesses influence the verification effort in the application domain of the analyzer?
\end{description}
It is worth noting that the performance improvement from technology-agnostic correctness witnesses is expected to be modest. In fact, \citet{Beyer2016} observed no consistent trend in performance gains.

\paragraph{Experimental Setup and Data.}
Our benchmarks are executed on a laptop running Ubuntu 22.04.3 on an \textsc{AMD Ryzen 7 PRO 4750U} processor.
For reliable measurements, all the experiments are carried out using the \textsc{BenchExec} framework~\cite{BenchExec}, where each tool execution is limited to \qty{900}{\second} of CPU time on one core and \qty{4}{\giga\byte} of RAM.
The benchmarks, tools and scripts used, as well as the raw results of the evaluation, are openly archived on Zenodo~\cite{zenodo-artifact}.

\subsubsection{Implementation.}
\textsc{Goblint} is an abstract interpretation framework for C programs~\cite{Vojdani2016}.
We have extended the framework with unassume operators and YAML witness support.
The correctness witnesses proposed by \citet{Beyer2016} and subsequently used in SV-COMP~\cite{Beyer2023} provide invariants using an automaton in the GraphML format.
The witnesses we consider (defined in \cref{sec:witnesses}) are much simpler and, thus, we use the newly-proposed YAML format~\cite{SoSyLab2021,TacasMeeting}, which directly matches our notion.
To this end, our implementation includes parsing of YAML witnesses and matching provided invariants to program locations such that the unassume operator can be applied.
Our implementation contains two unassume operators:
\begin{enumerate}
    \item Propagating unassume (\cref{sec:propagating-unassume}) for non-relational domains.
    The existing propagating assume in \textsc{Goblint} could be generalized and directly reused, yielding an unassume operator capable of handling, e.g., C lvalues, not just variables, with no extra effort (see \cref{app:pointers}).

    \item Strengthening-based dual-narrowing unassume (\cref{sec:dualnarrowing-unassume}) for relational domains.
    Although \textsc{Apron}~\cite{Jeannet2009}, which \textsc{Goblint} uses for its relational domains, does not provide dual-narrowing, the generic approach described in \cref{ex:unassume-standard-rel-fix} works for, e.g., octagons and convex polyhedra.
    Since the relational analysis is just numeric, $V$ is collected syntactically.
\end{enumerate}
To prevent unintended precision loss when widening from initially reached abstract values to the unassumed ones, we must take care to delay the application of widening.
We tag abstract values with the identifiers of incorporated witness invariants (UUIDs from the YAML witness) and delay the widening if this set increases~\cite{Mihaila_2013}.
Such widening tokens ensure that each witness invariant can be incorporated without immediate overshooting.

\subsection{Precision Evaluation}

We collected and provide a set of 11 example programs (excluding duplicates) from literature~\cite{Mine2017,Halbwachs_2012,Boutonnet_2017,Amato2013} where more advanced abstract interpretation techniques are developed to infer certain invariants, where standard accelerated solving strategies fail.
We configure \textsc{Goblint} the same as in SV-COMP~\cite{Saan2021,Saan2023}, except autotuning is disabled and relational analysis using polyhedra is unconditionally enabled.
We manually created YAML witnesses containing suitable loop invariants for these programs.
We also used two state-of-the-art verifiers from SV-COMP 2023 to generate real witnesses:
\textsc{CPAchecker}~\cite{Beyer2011,Dangl2015} and \textsc{UAutomizer}~\cite{Heizmann2013,Heizmann2023}.
Both verifiers are able to verify these programs and produce GraphML witnesses.
Following \citet{Beyer_2022a}, we use \textsc{CPAchecker} in its \texttt{witness2invariant} configuration to convert them into YAML witnesses that \textsc{Goblint} can consume.

\renewcommand{\s}{\cellcolor{green!30}\ding{51}}%
\newcommand{\e}{\cellcolor{red!30}\ding{55}}%
\newcommand{\w}{\cellcolor{yellow!30}{\ding{51}\ding{55}}}%
\newcommand{\invalid}{---}
\begin{table}
    \caption{Evaluation results on literature examples (excluding duplicates).
    The \textsc{Goblint} column indicates whether it can verify the program without any witness.
    Remaining columns indicate results with corresponding witnesses: witness validated (\colorbox{green!30}{\ding{51}}), program verified with witness-guidance but witness not validated (\colorbox{yellow!30}{\ding{51}\ding{55}}) or program not verified with witness-guidance (\colorbox{red!30}{\ding{55}}).}
    \label{tab:evaluation-precision}
    \centering
    \begin{tabular}{llc@{\hskip 1.5\defaultaddspace}ccc}
        \toprule
         & & \multirow{2}{1.7cm}[-0.1cm]{\centering \textsc{Goblint} w/o witness} & \multicolumn{3}{c}{\textsc{Goblint} w/ witness from} \\
        \cmidrule{4-6}
        Author(s) & Example & & Manual & \textsc{CPAchecker} & \textsc{UAutomizer} \\
        \midrule
        \citet{Mine2017} & 4.6 & \e & \s & \w & \s \\
         & 4.7 & \e & \s & \w & \s \\
         & 4.8 & \s & \s & \s & \s \\
         & 4.10 & \s & \s & \w & \s \\
        \addlinespace
        \multirow[t]{3}{2.3cm}{\citet{Halbwachs_2012}} & 1.b & \s & \s & \w & \w \\
         & 2.b & \e & \s & \w & \s \\
         & 3 & \e & \s & \w & \w \\
        \addlinespace
        \multirow[t]{3}{2.3cm}{\citet{Boutonnet_2017}} & 1 (polyhedra) & \e & \s & \w & \w \\
         & 3 & \e & \s & \e & \s \\
         & ``additional'' & \e & \s & \e & \s \\
        \addlinespace
        \parbox{1.7cm}{\citet{Amato2013}} & \texttt{hybrid} & \e & \s & \w & \s \\
        \midrule
        Total & 11 & \colorbox{green!30}{\ding{51}: 3} & \colorbox{green!30}{\ding{51}: 11} & \colorbox{yellow!30}{\ding{51}\ding{55}: 8}, \colorbox{green!30}{\ding{51}: 1} & \colorbox{yellow!30}{\ding{51}\ding{55}: 3}, \colorbox{green!30}{\ding{51}: 8} \\
        \bottomrule
    \end{tabular}
\end{table}

The results are summarized in \cref{tab:evaluation-precision}.
\textsc{Goblint} manages to verify the desired property for 3 of these programs without any witness,
but can validate all handwritten witnesses,
despite not implementing any of the advanced techniques needed for their inference.
With \textsc{CPAchecker} witnesses our validator can verify 9 out of 11 programs and validate 1 out of 11 witnesses.
With \textsc{UAutomizer} witnesses our validator can verify all 11 programs and validate 8 out of 11 witnesses.
Furthermore, our abstract interpreter can validate the witnesses from model checkers \emph{orders of magnitude} faster than it took to generate them.

The evaluation, however, shows many instances where the program was only verified thanks to witness-guidance, but not all witness invariants could be validated, especially for \textsc{CPAchecker}.
This is precisely due to the phenomenon described in \cref{ex:validity-concrete-vs-abstract}: in these small programs bounded model checking is successful and yields disjunctive invariants over all the finitely-many cases.
Surprisingly, an invariant is useful for witness-guided verification, even when it cannot be proven to hold abstractly.
\begin{example}
\label{ex:unassume-disjunctive}
The invariant from \cref{ex:validity-concrete-vs-abstract} relaxes an abstract state:
\[
    \sems{\unassume_{\{\texttt{x}\}}(\texttt{x} = 1 \lor \texttt{x} = 2)}\,\{\texttt{x} \mapsto [1, 1]\} = \{\texttt{x} \mapsto [1, 2]\}.
\]
\end{example}

\subsection{Performance Evaluation}

To explore whether a suitable witness can reduce verification effort,
we consider larger programs, as runtimes for the literature examples are negligible.
Since \textsc{Goblint} specializes in the analysis of multi-threaded programs,
we examine a set of multi-threaded POSIX programs
previously used to evaluate \textsc{Goblint}~\cite{Schwarz_2021,SchwarzSSEV23}.
We manually construct witnesses that contain core invariants for these programs,
based on how widenings were applied during fixpoint solving.
We configure \textsc{Goblint} as described earlier, but with relational analysis disabled.
In addition to CPU time, we measure analysis effort without a witness and with witness-guidance via transfer function evaluation counts.
This metric of evaluations is proportional to CPU time, but excludes irrelevant pre- and post-processing, and is independent of hardware.

\begin{table}
    \caption{Evaluation results on \textsc{Goblint} benchmarks.
    The LLoC column counts logical lines of code, i.e., only lines with executable code, excluding declarations.}
    \label{tab:evaluation-performance}
    \centering
    \begin{tabular}{lS[table-format=4.0]@{\hskip 1.5\defaultaddspace}S[table-format=5.0]S[table-format=2.2]@{\hskip 1.5\defaultaddspace}S[table-format=5.0]S[table-format=2.2]@{\hskip 1.5\defaultaddspace}S[table-format=2.1]S[table-format=2.1]}
        \toprule
         &  & \multicolumn{2}{c}{w/o witness} & \multicolumn{2}{c}{w/ witness} & \multicolumn{2}{c}{Reduction} \\
        \cmidrule(l{0.1em}r{0.7em}){3-4}\cmidrule(l{0.1em}r{0.7em}){5-6}\cmidrule{7-8}
        Program & LLoC & {Evals} & {CPU time (s)} & {Evals} & {CPU time (s)} & {Evals} & {CPU time} \\
        \midrule
        pfscan & 559 & 4194 & .86 & 2919 & .73 & 30.4\percent & 15.4\percent \\
        aget & 587 & 7932 & 2.23 & 4683 & 1.68 & 41.0\percent & 24.7\percent \\
        knot & 981 & 29588 & 4.92 & 21432 & 4.54 & 27.6\percent & 7.7\percent \\
        smtprc & 3037 & 48559 & 15.00 & 24091 & 7.95 & 50.4\percent & 47.0\percent \\
        \midrule
        Average &  &  &  &  &  &  37.3\percent & 23.7\percent \\
        \bottomrule
    \end{tabular}
\end{table}

The results, aggregated in \cref{tab:evaluation-performance}, show a noticeable performance improvement in the abstract interpreter when guided by a witness.
However, the fixpoint-solving process still requires numerous widening iterations.
This is due to various abstractions used by \textsc{Goblint} that cannot be expressed as C expressions, including but not limited to array index ranges in abstract addresses and various concurrency aspects.
Nevertheless, the average 1.23$\times$ CPU time speedup is relatively close to the average 1.63$\times$ improvement achieved by \citet{Albert_2005} when using analyzer-specific certificates (see \cref{sec:related-work}).

Admittedly, we have used a limited set of benchmarks and hand-crafted witnesses
because our automatically generated witnesses produce excessive information.
Large witnesses that express full proofs with numerous invariants can be problematic for a validator~\cite{Beyer2016,Albert_2006}, which must manipulate, use, and/or verify them.
In our case, the validator performs an unassume operation for each invariant each time the corresponding transfer function is evaluated.
The speedup gained from using the witness must outweigh the overhead to truly benefit from witnesses.
This amounts to witnesses containing partial proofs, like loop invariants~\cite{Beyer2016}.
Moreover, our approach does not take advantage of exact invariants, such as equalities outside of disjunctions,
since these do not relax a reached state already representing such exact values.
Even if such invariants are useful for some validators, they do not benefit our witness-guided verifier.
Therefore, the challenge remains for us, in collaboration with other tool developers,
to develop methods for generating suitable witnesses.

\section{Related work}
\label{sec:related-work}
Fixpoint iterations involving widening and narrowing are well-studied~\cite{CousotC77,Amato2013,Amato2016,Seidl_2021,10.1007/978-3-662-46081-8_2}, but focus mostly on improving precision and ensuring termination.
\citet{Halbwachs_2012} extend fixpoint iteration with partial restarting, which derives from the narrowing result a new initial value for the following widening iteration, hoping it improves the result.
Their restarted value is analogous to our partial initialization and could be used as such.
They focus on finding such values automatically, while we focus on using them to avoid all the computation leading up to it.
\citet{Boutonnet_2017} improve the technique for finding good restarting candidates.
\citet{10.1007/978-3-662-46081-8_2} extends fixpoint iteration with dual-narrowing, hoping it improves the result further.
Both approaches focus purely on improving precision with more iterations, while we aim to skip that iteration and arrive at the same result quicker, knowing the invariant.
Hence, the techniques can be combined: use theirs to find a precise invariant and use ours to directly reuse it.

\citet{Arceri2022} swap the abstract domain for a more precise one when switching from widening to narrowing.
This can be considered a precision improvement technique, which makes the narrowing phase more expensive.
However, it can also be viewed as an optimization, which makes the widening phase cheaper.
Either way, it can be combined with our approach: immediately using the more precise domain with the final invariant, ideally skipping iteration in both ways.

Widening operators themselves are also well-studied~\cite{Blanchet_2003,Bagnara_2005,Gopan_2006,Mihaila_2013,Mine2017}.
Widening \emph{up to} or \emph{with threshold} use candidate invariants as intermediates to avoid irrecoverably losing precision.
Such automated techniques can identify which candidates are true invariants.
Using these as input to our approach is an effective way of supplying known good thresholds at specific program points, removing the need for retrying all the candidates on re-verification.
This is one instance of dealing with the inherent non-monotonicity of widening operators~\cite{10.1007/978-3-662-46081-8_2}.
Furthermore, widening thresholds require domain-specific implementation, whereas our approach is more generic.

Our na\"ive unassume with its havoc and assume bears some similarity to the generation of  \emph{verification conditions} from user-supplied loop invariants~\cite{FlanaganSaxe01}.
However, there one additionally refines the state with the loop condition and then checks that the loop invariant is preserved.
We avoid the former to remain sound by construction, but effectively do the latter when validating witness invariants.

\citet{Albert_2005} introduce \emph{Abstraction-Carrying Code} (ACC) as an abstract-interpretation--based instance of Proof-Carrying Code (PCC).
For validation they use a simplified analyzer which only performs a single pass of abstract interpretation and no fixpoint iteration.
Thus, it requires certificates to supply invariants for all loops.
This is more restrictive than our validation approach, which runs in a single pass if all necessary invariants are provided, but also allows some fixpoint iteration if this is not the case.
Nevertheless, we could handicap our analyzer with this stronger restriction.

\citet{Albert_2006} develop a notion of reduced certificates which can be smaller and are used by the validator to reconstruct full certificates.
\citet{Besson_2006} propose a fixpoint compression algorithm to further compact the certificates.
In follow-up work, \citet{Besson_2007} develop a theory for studying the issue of certificate size.
Rather than using the strongest information from the least fixpoint of an analysis, they seek the weakest information still sufficient for implying correctness.
This omits irrelevant information, leading to smaller witnesses.
Such techniques could also be used when generating witnesses for our validator.

\section{Conclusion}
We have demonstrated how to turn abstract-interpretation--based tools into witness-guided verifiers and witness validators, by equipping them with unassume operations.
These can be constructed from abstract transformers for assumes, non-deterministic assignments, joins and (optionally) dual-narrowings, which allow retaining more precision for relational abstract interpretation.
A powerful syntax-directed unassume operation for non-relational domains can be derived from a classical algorithm with minimal changes.
Our implementation and evaluation demonstrate that unassuming invariants from witnesses can both speed up the analysis and make it more precise.
The experiments further show that the abstract interpreter can benefit from witnesses produced by model checkers,
and thus indicate that the approach is suited even for cross-technology collaboration.

\paragraph{Acknowledgements.}
This work was supported by Deutsche Forschungsgemeinschaft (DFG) – 378803395/2428 \textsc{ConVeY} and Shota Rustaveli National Science Foundation of Georgia under the project FR-21-7973.

\clearpage

\clearpage % flush floats before references

\bibliography{thread-witnesses}

\newpage
\appendix

\section{Pointers}
\label{app:pointers}
In \cref{sec:propagating-unassume} we claim that the algorithms generalize from variables to lvalues.
The key machinery for this is the \emph{variable set variance} described in the paper.
While the original HC4-revise algorithm directly refines variables at leaf nodes,
we handle pointers by letting leaf nodes create partial states containing the dynamically resolved variables.

Let $l ::= x \;|\; \texttt{*}x$ be the grammar for lvalues, which can either be variables or dereferencing of variables.
Let abstract value domain for pointer variables be may-point-to sets of variables $2^{\mathcal{V}}$.
Let constants include address-taking as $\texttt{\&}x$ for a variable $x \in \mathcal{V}$.

\paragraph{Evaluation.}
The forward evaluation of address-taking is defined by ${(\texttt{\&}x)^\sharp = \{x\}}$.
Forward evaluation of lvalues uses the following helper function for evaluating lvalues to may-point-to sets:
\begin{align*}
    &\sems[\mathbb{L}]{l} : \mathbb{D} \to 2^{\mathcal{V}} \\
    &\sems[\mathbb{L}]{x}\,d = \{x\} \\
    &\sems[\mathbb{L}]{\texttt{*}x}\,d = d\,x
\end{align*}
The forward evaluation of lvalues is then a join over all the possibilities:
\[
    \sems[\mathbb{E}]{l}\,d = \bigsqcup_{\mathclap{x \in \sems[\mathbb{L}]{l}\,d}} d\,x.
\]
\paragraph{Assume.}
The backward evaluation of lvalues is defined analogously:
\[
    \sems[\overleftarrow{\mathbb{E}}]{l}\,v'\,d = \bigsqcup_{\mathclap{x \in \sems[\mathbb{L}]{l}\,d}} d[x \mapsto d\,x \sqcap v'].
\]
However, it is easier understood using the following equivalent formulation:
\[
    \sems[\overleftarrow{\mathbb{E}}]{l}\,v'\,d = \begin{cases}
        d[x \mapsto d\,x \sqcap v'], &\text{if } \sems[\mathbb{L}]{l}\,d = \{x\}, \\
        d, &\text{otherwise}.
    \end{cases}
\]
Thus, in the case of an ambiguous may-point-to set, no refinement takes place.
This is because each variable is refined in one of the joinees, while left untouched in others, leaving the joined value also unchanged.

\paragraph{Unassume.}
Unassuming of lvalues is also defined analogously:
\[
    \sems[\tilde{\mathbb{E}}]{l}\,v'\,d = \bigsqcupstar_{\mathclap{x \in \sems[\mathbb{L}]{l}\,d}} \{x \mapsto v'\}.
\]
However, it is also easier understood using the following equivalent formulation:
\[
    \sems[\tilde{\mathbb{E}}]{l}\,v'\,d = \{x \mapsto v' \mid x \in \sems[\mathbb{L}]{l}\,d\}.
\]
This can be seen as a special case of abstract assignment:
\[
    \sems[\tilde{\mathbb{E}}]{l}\,v'\,d = \sems{l \gets \gamma\,v'}\,\{x \mapsto \bot \mid x \in \sems[\mathbb{L}]{l}\,d\},
\]
where $\sems{l \gets e}\,d$ is the abstract operator for assigning the value of expression $e$ to the lvalue $l$.
In this special case the assigned abstract value is $v'$, not $\sems[\mathbb{E}]{e}\,d$.

Therefore, propagating unassume for lvalues can be implemented using existing primitives: backward operators and abstract assignment.
This is not limited to just simple pointers but works the same way for C lvalues, which also include array indexing and structure field offsets.

\end{document}